\documentclass[twocolumn,pra,nobalancelastpage,aps,raggedbottom]{revtex4-2}
\usepackage{times}

\usepackage{amsmath,amsfonts,amssymb,graphicx,hyperref,epstopdf,xcolor,tikz,scalerel,natbib}
\usepackage{mathptmx,textcomp,braket}
\usepackage[T1]{fontenc}
\usepackage[utf8]{inputenc}

\definecolor{lime}{HTML}{A6CE39}
\DeclareRobustCommand{\orcidicon}
{
	\begin{tikzpicture} 
	\draw[lime, fill=lime] (0,0) circle [radius=0.15] node[white] {{\fontfamily{qag}\selectfont \tiny ID}};
	\draw[white, fill=white] (-0.0625,0.095) 	circle [radius=0.007];
	\end{tikzpicture}
	\hspace{-2.2mm}
}
\newcommand\orcidID[1]{\href{https://orcid.org/#1}{\orcidicon}}

\newcommand{\pddt}[1]{\partial#1/\partial t}

\newcommand{\be}{\begin {equation}}
\newcommand{\ee}{\end {equation}}
\newcommand{\beqa}{\begin {eqnarray}}
\newcommand{\eeqa}{\end {eqnarray}}
\newcommand{\mb}{\mathbf}
\newcommand{\Sch}{Schr\"odinger }
\newcommand{\Exp}[1]{\text{e}^{#1}}
\hypersetup{colorlinks,citecolor=blue,filecolor=black,linkcolor=blue,urlcolor=blue}

\begin{document}

\title{Controlling resonant enhancement in higher-order harmonic generation}

\author{Rambabu Rajpoot\orcidID{0000-0002-2196-6133}}

\author{Amol R. Holkundkar\orcidID{0000-0003-3889-0910}}
\email[E-mail: ]{amol.holkundkar@pilani.bits-pilani.ac.in}
 
\author{Jayendra N. Bandyopadhyay\orcidID{0000-0002-0825-9370}}

\affiliation{Department of Physics, Birla Institute of Technology and Science - Pilani, Rajasthan,
333031, India}

\date{\today}

\begin{abstract}
We present a method to tune the resonantly enhanced harmonic emission from engineered potentials, which would be experimentally feasible in the purview of the recent advances in atomic and condensed matter physics. The recombination of the electron from the potential dependent excited state to the ground state causes the emission of photons with a specific energy. The energy of the emitted photons can be controlled by appropriately tweaking the potential parameters. The resonant enhancement in high-harmonic generation enables the emission of very intense extreme ultra-violet or soft x-ray radiations. The scaling law of the resonant harmonic emission with the model parameter of the potential is also obtained by numerically solving the time-dependent Schr\"odinger equation in two dimensions.
\end{abstract}

\maketitle

\section{Introduction}

The rapid advancement in the field of higher-harmonic generation (HHG) led to the development of the extreme ultraviolet (XUV) and soft x-ray radiation sources \cite{Mairesse2003_Science,Liu2019SpecLett}, which indeed are crucial for the generation of the electromagnetic pulses at attosecond time scale \cite{Hentschel2001_Nature,Krausz2009_RMP,Corkum2007_NatPhy}. These attosecond pulses promise variety of the applications, for example the study of electron correlation effect, detailed microscopic motion of electrons in atoms or molecules \cite{Chini2014_nat,Krausz2009_RMP}, observing the temporal delays in photoemission from atomic orbitals \cite{PhysRevLett.115.153001}, ultrafast chiral-specific dynamics in molecules \cite{Ferre2015}, x-ray magnetic circular dichroism spectroscopy \cite{Kfir2015} to name a few. The generation of the higher-harmonics from the laser-atom interaction can be understood by a celebrated \lq\lq three-step model\rq\rq \cite{Corkum1993_PRL}, which was a semi-classical model wherein the higher harmonic generation is described as a three-step process, ionization of the electron, free propagation of electron in laser field and finally the recombination of the electron with the parent ion. Additional kinetic energy is emitted in the form of the higher harmonics of the fundamental frequency of the interacting laser pulse.  

After the conceptualization of the HHG, the research in this field is aimed toward enhancing the harmonic cutoff of the HHG and increasing the corresponding intensity of the emitted harmonics. Different pulse shaping techniques are introduced in the past to achieve these objectives \cite{PhysRevA.97.053414, PhysRevLett.117.093003,PhysRevA.96.033407,PhysRevA.93.033404,Rajpoot_2020}. However, the resonant enhancement of harmonic emission is now seen to be a very promising aspect of HHG \cite{Toma_1999,PhysRevA.65.023404}. These resonantly enhanced harmonics lead to an increase in the attopulse intensity and relax the requirements for XUV filtering from emitted spectra \cite{PhysRevA.94.063420}. On similar grounds, the presence of the Giant Autoionization Resonance in the harmonic emission of the manganese laser-ablated plasma plume is recently reported \cite{PhysRevLett.121.023201}. The multiphoton resonance \cite{PhysRevA.78.053406}, the Fano resonance in HHG under the purview  of strong field approximation \cite{PhysRevA.89.053833}, the shape resonances in the HHG \cite{PhysRevA.84.013430}, resonances due to multielectron effects in HHG \cite{PhysRevLett.118.203202}, resonances in \textit{ortho} and \textit{para} Helium \cite{Shi_2021} are examples of some more observed phenomena. Moreover, the presence of the resonant harmonic can also be understood by the formation of the auto-ionizing energy states in the continuum \cite{Fareed2017,PhysRevLett.104.123901}. The experiments on the harmonic emission through the plasma plumes demonstrated the resonant harmonic enhancement \cite{PhysRevA.74.063824,PhysRevA.75.063806,PhysRevLett.102.013903}. Previously various model for the resonant harmonic enhancements based on the bound-bound transitions \cite{PhysRevA.65.023404}, bound-continuum transitions, stimulated emission through bound and autoionizing states \cite{Milo_evi__2007} etc are proposed,  however the 4 step model proposed by the Strelkov \cite{PhysRevLett.104.123901} was able to demonstrate the physics aspects of the resonant enhancement of harmonic emission reported in the experiments. 

This resonant enhancement in the HHG not only shed light on the fundamental processes of the HHG but also enabled the researchers to remove the need for any filtering and result in very intense attosecond pulses. In this work, we aim to propose a method wherein engineered potentials can tweak the location of the resonant harmonic. The engineering of new materials and their role in the HHG are now widely explored for improved harmonic generation efficiency \cite{PhysRevA.103.043117,Tancogne-Dejeaneaao5207}. Given the recent advances in the field of atomic and condensed matter physics, in the near future, it would be feasible to modify the potential of the atom participating in the HHG, either by applying some external fields or by the presence of an atom in the vicinity \cite{Shi_2021}. The tuning of the resonantly enhanced harmonic would enable the generation of very high intense attosecond pulses in the XUV/x-ray regime of electromagnetic spectrum.    
 
The paper is organized as follows. First, details of the simulations are discussed in Sec. \ref{sec2}, followed by the results and the discussions in Sec. \ref{sec3} and finally, the concluding remarks and future directions are discussed in Sec. \ref{sec4}. 

\section{Numerical Methods}
\label{sec2} 

We study the interaction of the laser with the atomic system with given model potential using 2D Time Dependent 
\Sch Equation (TDSE) solver in the Cartesian grid. The atomic units are used throughout the manuscript unless otherwise stated, and hence $e = m_e = \hbar = 1$. The TDSE under dipole approximation in the length gauge (LG) is written as, $i \pddt{\psi(\mb{r},t)}   = H(t)\psi(\mb{r},t)$, where, $H(t)$ is the time-dependent Hamiltonian of the system given by,
\be H(t) = - \frac{1}{2} \nabla^2\ +\ V(r,a_0)\ +\ \mb{r}\cdot\mb{E}(t,\epsilon).\ee  
The time dependent electric field amplitude of laser pulse is given by:
\be \mb{E}(t,\epsilon) = F_0 \\ 
              \begin{cases}
                \sin^2(\pi t/T) \cos(\omega_0 t) \mb{e}_{x},      &\varepsilon = 1\\
                [0.05+\sin^2(\pi t/T) \cos(\omega_0 t)]\mb{e}_{x},      &\varepsilon = 2\\
                \exp\Big[\frac{-4\ln(2) (t-\eta)^2}{T^2}\Big] \cos[\omega_0 (t-\eta)] \mb{e}_x,  &\varepsilon = 3
              \end{cases}
\ee
where, $F_0\ \text{[a.u.]} \sim 5.342\times 10^{-9} \sqrt{I_0}$, is the field amplitude of the electric field in atomic units with $I_0$ being the laser intensity in units of $\text{W cm}^{-2}$, $\omega_0 \sim 0.057$ a.u. is the frequency of the laser (800 nm wavelength), $T$ is the pulse duration (total or FWHM) of the laser, $\eta$ is some phase and $\epsilon$ is just a flag to select a particular laser envelope. The potential function in the Hamiltonian is expressed as:
\be V(r,a_0) = -\frac{1 + 9 \exp[-r^2]}{\sqrt{r^2 + a_0}} \label{poten}\ee
here, $r =  \sqrt{x^2+y^2}$ and $a_0$ is the soft-core potential parameter. The soft-core potential parameter $a_0$ decides the bound state energies of the model potential and thus it controls the atomic system(s) under study. This kind of potential function is routinely used in the context of the Neon atom under single active electron configuration \cite{Zhang2020_PRA,Zhang2018_PRA,Lukas2015_PRA}. The energies of the bound states for a given soft-core potential parameter $a_0$ are accurately estimated by the spectral method as proposed by Fiet et al \cite{Feit1982_JCP}.

\begin{figure}[b]
\includegraphics[totalheight=0.48\columnwidth]{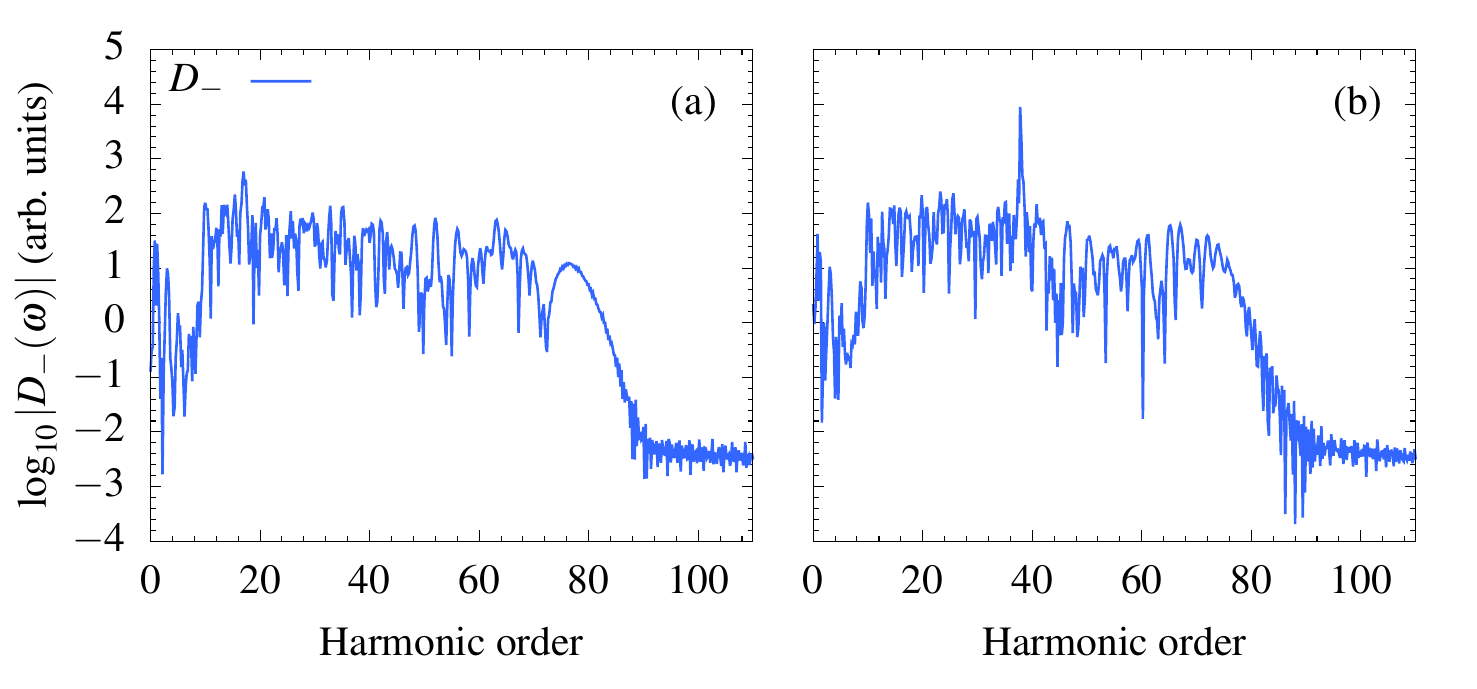}
 \caption{Intensity of left  rotating harmonics is presented. The spectrum is obatained after the interaction of the 800 nm laser, having $T = 5T_0$, and $I_0 = 5\times 10^{14}$ W cm$^{-2}$ with a current carrying state $\ket{\phi_{+}}$ state of Neon atom. The laser pulse envelope $\mb{E}(t,1)$ (a) and $\mb{E}(t,2)$ (b) are used respectively.}
 \label{fig1} 
\end{figure}
 
The general solution of the TDSE is obtained by employing the time evolution operator $U(t_0+\Delta t, t_0)$ on the  initial state wave-function $\psi_0(\mb{r},t_0)$ of the electron,   
\be\psi(\mb{r},t_0+\Delta t) = U(t_0+\Delta t, t_0) \psi_0(\mb{r},t_0)\ee
and the TDSE is solved numerically by employing the split-operator method \cite{Feit1982_JCP}. Wherein the time evolution operator is factored as a product of the exponential of kinetic and potential energy operators, as:
\be U(t_0+\Delta t, t_0) \simeq \Exp{-i\mb{p}^2\Delta t/4} \Exp{-iV_\text{eff}(t_0+\Delta t/2)\Delta t} \Exp{-i\mb{p}^2\Delta t/4}\ee
where, $\mb{p^2} \equiv p_x^2 + p_y^2$, $V_\text{eff}(t) = V(r,a_0) + \mb{r}\cdot\mb{E}(t,\epsilon)$ is the effective potential in the length gauge, and $\Delta t$ is the simulation time step. The initial state $\psi_0(\mb{r},t_0)$ is obtained using the imaginary-time propagation method \cite{Bader2013_JChemPhys}. The well known Gram-Schmidt orthogonalization process is employed to project out the ground state to obtain the $p_x$ and $p_y$ orbitals. In order to avoid nonphysical reflections at the spatial grid boundaries a mask function,
\be M_{abs}(x,y) = \frac{1}{[1 + \exp(\kappa |x| - x_\text{abs})] [1 + \exp(\kappa |y| - y_\text{abs})] }, \ee
with the damping parameter $\kappa = 1.2$ is introduced. In our calculations, the spatial domain has the maximal extent of $\pm 150$ a.u. and last 10 a.u. distance is used for the masking purpose, which implies $x_\text{abs} = y_\text{abs} \sim 142$ a.u. The spatial grid mesh has $2048 \times 2048$ points i.e. spatial step size $\Delta x = \Delta y \sim 0.146$ a.u. and the simulation time step $\Delta t \sim 0.008$ a.u. is considered. The convergence is tested with respect to the spatial grid as well as space and time steps. The time-dependent dipole acceleration is obtained through the Ehrenfest theorem as \cite{sandPRL_1999}:
\be \mb{a}(t) = - \langle \psi(\mb{r},t) | \nabla V(r,a_0)\ +\ \mb{E}(t,\epsilon) | \psi(\mb{r},t) \rangle\ee
and the harmonic spectra is then obtained by taking the Fourier transform of the dipole acceleration, separately along $x$ and $y$ direction,
\be S_{x,y}(\omega) =  \frac{1}{\sqrt{2\pi}} \int a_{x,y}(t) \exp[-i\omega t]\ dt. \ee 
We can also construct the left ($S_{-}$) and right ($S_{+}$) rotating harmonic components by taking the linear combinations of $S_x(\omega)$ and $S_y(\omega)$ as: $S_{\pm} = (S_x \pm i S_y)/\sqrt{2}$. The intensities of the left and the right rotating harmonic components are given by $D_{\pm} = |S_{\pm}|^2$, and hence the ellipticity of the harmonic radiation is calculated by $\delta = (|S_{+}| - |S_{-}|)/(|S_{+}| + |S_{-}|)$ \cite{Zhang2020_PRA,Zhang2018_PRA}.  It is observed in these work that the typical characteristics of the HHG spectra are similar for both left and right circularly polarized harmonics, and hence throughout the manuscript we present the left circularly polarized spectrum ($D_-$) for no particular reason.  

In the following, we present the results pertaining to the effect of the laser temporal envelope on the harmonic emission and the presence of the resonance peak. We then later will comment on the aspects related to controlling the same by tweaking the model potential.

\section{Results and discussions}
\label{sec3}

As we have mentioned in the previous section that the initial state $\braket{\mb{r}|\psi_0}$ is obtained using the imaginary-time propagation method, and then the Gram-Schmidt orthogonalization method is used to project out the ground state thus obtained $p_x$ and $p_y$ orbitals, say $\ket{\phi_{p_x}}$ and $\ket{\phi_{p_y}}$ respectively. We can also consider a linear combination of the $\ket{\phi_{p_x}}$ and $\ket{\phi_{p_y}}$ to be considered as an initial state, i.e. $\ket{\psi_0} \equiv \ket{\phi_{\pm}} = [\ket{\phi_{p_x}} \pm i \ket{\phi_{p_y}}]/\sqrt{2}$ \cite{PhysRevLett.115.153001}. The states $\ket{\phi_{\pm}}$ are referred as the current-carrying states, which are recently shown to be capable of controlling the ellipticity of the generated isolated attosecond pulse \cite{Zhang2020_PRA}.

\begin{figure}[t]
\includegraphics[totalheight=0.65\columnwidth]{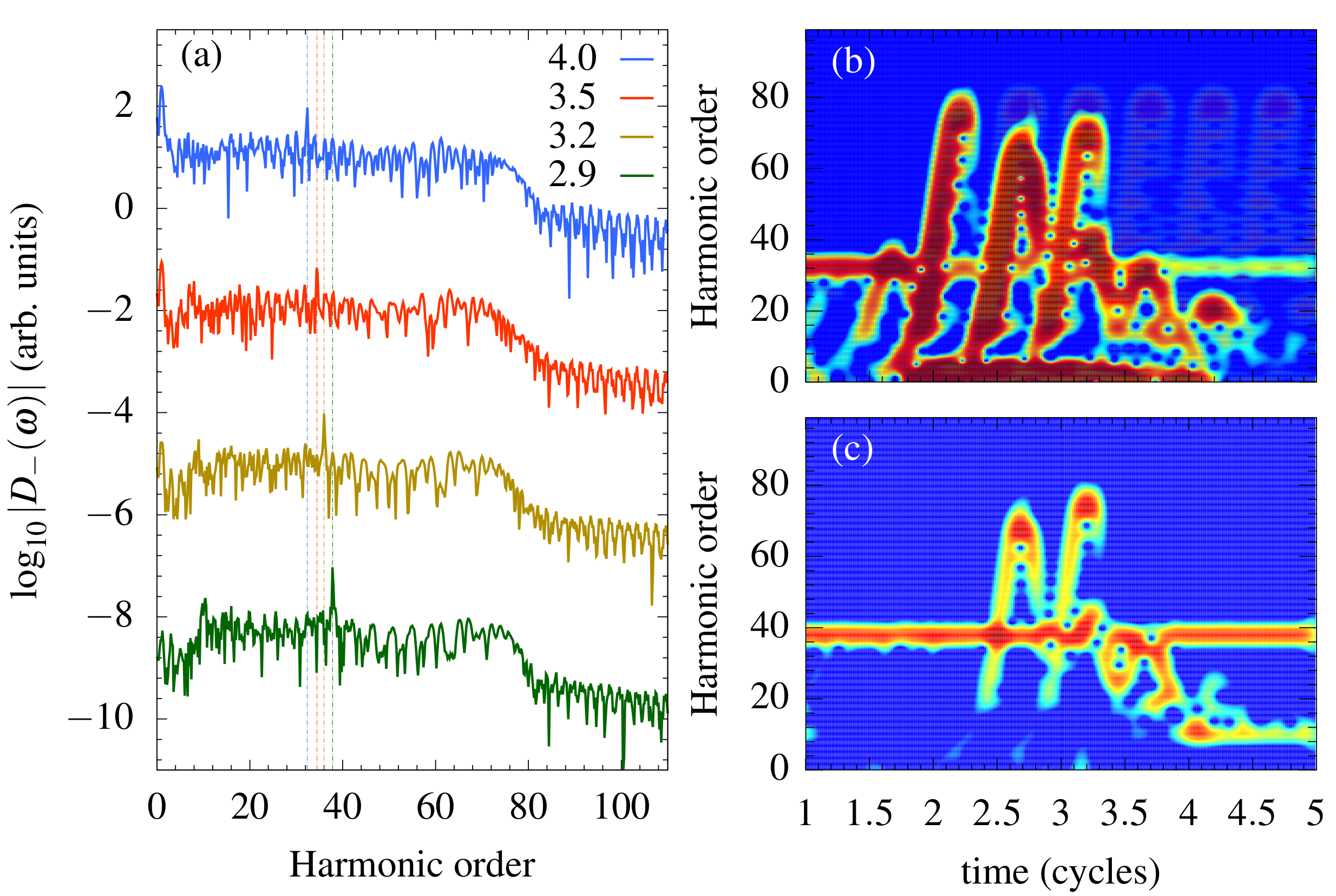}
 \caption{Higher harmonic spectra for different values of the potential parameter $a_0$ are presented (a). Please note that all the curves in (a) are shifted equally for visual appeal. The time-frequency Gabor transform for $a_0 = 4$ (b) and $a_0 = 2.9$ (c) are also presented. The $\mb{E}(t,2)$ laser pulse is used with peak intensity $I_0 = 5\times 10^{14}$ W cm$^{-2}$ and pulse duration $T = 5T_0$.}
 \label{fig2} 
\end{figure}
   
In Fig. \ref{fig1} we present the intensity of the left rotating harmonic component of the HHG spectra obtained by the laser interaction with the Neon atom. The soft-core potential parameter $a_0 = 2.88$ is considered so that the energy of the initial state $\ket{\phi_{+}}$ is found to be $E_{+} \sim 0.79$ a.u., which is close to the experimentally obtained ionization potential of the Neon atom. The wavelength of the laser is 800 nm, pulse duration $T = 5T_0$ [$T_0$ is an optical cycle], and the peak intensity is $I_0 = 5\times 10^{14}$ W cm$^{-2}$. Initially, we considered the $\mb{E}(t,1)$ field envelope, and the harmonic spectra are presented in Fig. \ref{fig1}(a). 

In Fig. \ref{fig1}(b) we kept all the laser and atomic parameters same as Fig. \ref{fig1}(a) but now considered the envelope $\mb{E}(t,2)$, i.e., a dc field of amplitude $5\%$ of peak laser electric field amplitude is applied along the same direction as the laser polarization, as a result a resonance peak at $\sim 37^\text{th}$ harmonic is observed. This resonance peak can be understood through the re-collisions of the participating electron in the given atomic potential \cite{PhysRevA.68.033403}. Using the spectral method \cite{Feit1982_JCP} the ground state   and the first excited energy of the potential $V(r,a_0)$ for the soft-core potential parameter $a_0 = 2.88$ are found to be $E_0 \sim -2.92$ a.u. and $E_1 \sim -0.79$ a.u. respectively. The current carrying state $\ket{\phi_+}$ is a linear combination of the $\ket{\phi_{p_x}}$ and $\ket{\phi_{p_y}}$ and hence this state will have  the same bound state energy $E_{+} \sim -0.79$ a.u. The dc electric field enables the transition of the re-colliding electron from the $\ket{\phi_+}$ state to the ground state $\ket{\phi_0}$ which results in the emission of the harmonic $\Delta E/\omega_0 = (E_+ - E_0)/\omega_0 \sim 37$, as seen in the Fig. \ref{fig1}(b). 

\begin{figure}[b]
\includegraphics[totalheight=0.7\columnwidth]{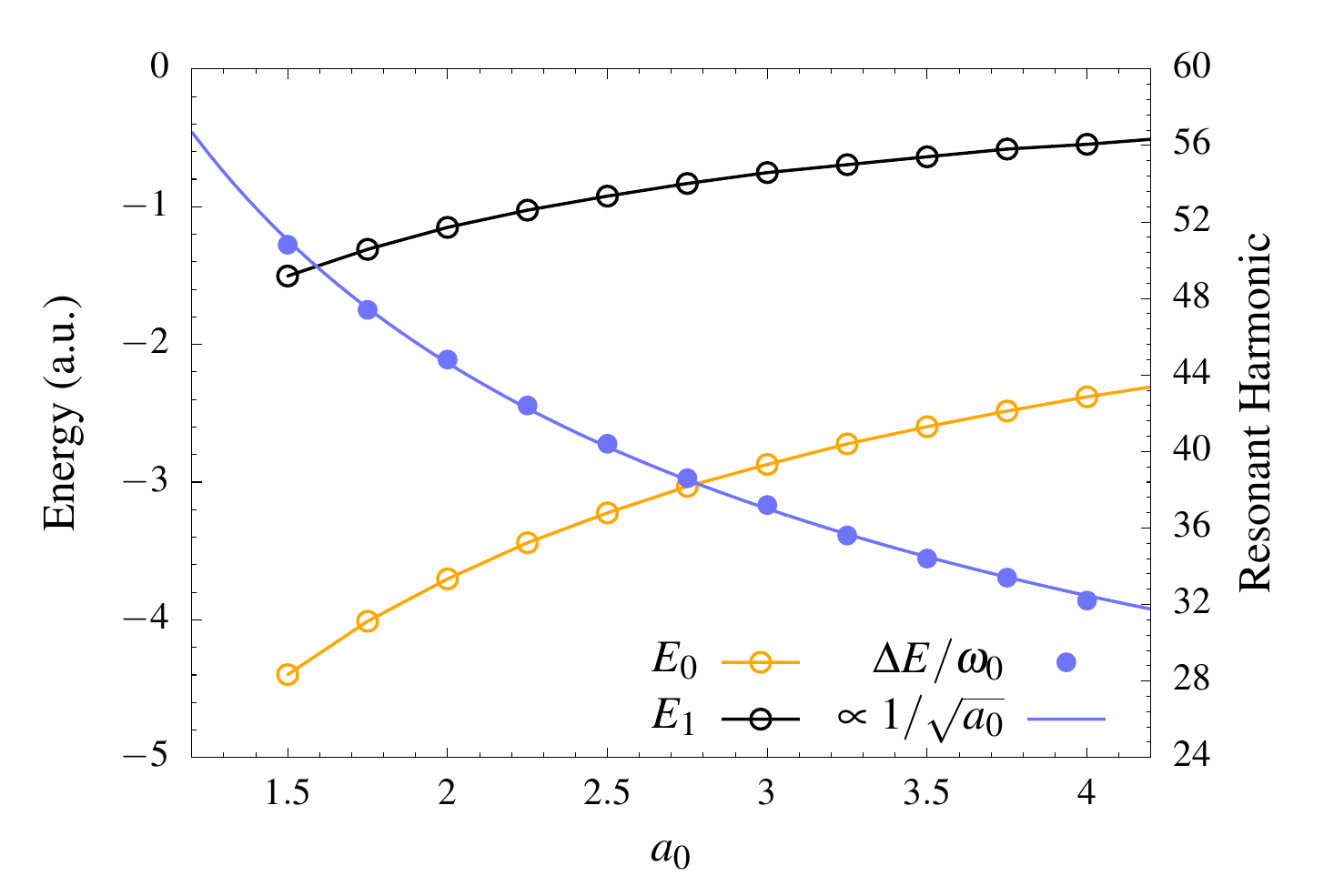}
 \caption{The variation of the ground state energy ($E_0$) and the first excited energy ($E_1$) with the potential parameter is presented. These bound energies are calculated by the spectral method. The resonant harmonic emission corresponds to the transition from excited state to the ground state is represented by blue solid circles on the right $y$ axis. The $\propto 1/\sqrt{a_0}$ scaling of the resonant harmonic is also shown by the solid line. }
 \label{fig3} 
\end{figure}

The harmonic emission in the presence of the shape resonances are previously reported \cite{PhysRevA.84.013430}, wherein a resonance peak is observed when an electron makes the transition from the metastable state to the ground state under the influence of the potential, which supports the metastable states in the electron continuum. These shape resonances are in accordance with the \textit{four} step model proposed by the Strelkov \cite{PhysRevLett.104.123901}, where the electron is captured in the metastable state and then finally recombine to the ground state. Furthermore, according to a very recently proposed idea, the recollision excitation gives rise to the resonance peaks in the para- and ortho-Helium \cite{Shi_2021}. Similarly, in the multielectron configuration, the cation can affect the recollision dynamics of the electron, giving rise the resonance peak in the harmonic spectra \cite{PhysRevA.99.043401,PhysRevLett.118.203202}. In both the cases of the ortho/para Helium \cite{Shi_2021}, and the multielectron configuration \cite{PhysRevLett.118.203202}, the recolliding electron dynamics is altered by the coloumbic potential caused by other electrons, which manifests in the transition from the excited state (metastable state) to the ground state emitting the harmonic corresponding to the energy difference. In the context of Fig. \ref{fig1}(b) this can be understood as follows. We are propagating a current-carrying state $\ket{\phi_+}$, the electron tunnel ionizes and accelerated freely in the continuum gaining energy, the electron upon return recollides with the state it started with giving rise to the harmonic peak near $\sim 14^\text{th}$ harmonic which is equivalent to the energy of the excited state [i.e. $\sim 0.79$  a.u.] it started with. The constant dc field regulates the population of the excited state, with which the recolliding electron interacts and causes the transition from the excited state to the ground state emitting the $\sim 37^\text{th}$ harmonic. In the absence of the dc field, the harmonic spectra are similar to those predicted under the SAE model, and no resonance peak in the HHG is observed. These aspects are previously corroborated under the multielectron configuration  \cite{PhysRevA.99.043401}. The resonance peaks in the multielectron configuration rely on the residual coulombic field, which triggers the dynamical electron correlation effect \cite{PhysRevLett.118.203202}. 
 
The resonance enhancement observed in Fig. \ref{fig1}(b) is the property of the potential function as it is caused by the transition from the bound states of the model potential. This aspect of the harmonic spectra enabled us to explore the effect of the soft-core potential parameter on the resonance enhancement in HHG. As we discussed, the resonance peak corresponds to the electronic transition from the first excited state to the ground state; if this fact is correct, this should result in the shift of the resonance peak with the variation of the potential parameter. To test this argument, we present in Fig. \ref{fig2}(a) the harmonic spectra for the different potential parameters. We consider the laser envelope $\mb{E}(t,2)$ with peak intensity $I_0 = 5\times 10^{14}$ W cm$^{-2}$ and duration $T = 5T_0$. The change in the potential parameter $a_0$ translates in the variation of the ground and excited states energies, and hence a clear shift in the resonance peak is observed. Furthermore, the time-frequency analysis of the emitted spectra can be studied using the Gabor wavelet transform. The Gabor transform of the emitted harmonic spectra for $a_0 = 4.0$ and $a_0 = 2.9$ are respectively presented in the Fig. \ref{fig2}(b) and Fig. \ref{fig2}(c). It can be seen that the emission of the resonant harmonic starts very early, giving rise to the enhanced emission of the same. Moreover, it can be observed from the Fig. \ref{fig2}(b) and (c) that on the same scale, the harmonic intensity for $a_0 = 2.9$ case is lower than the $a_0 = 4$ case. This is because the parameter $a_0$ relates to the typical distance from the nucleus, lower values of the same would result in the larger ionization threshold or tightly bounded electronic configuration. For the same laser parameter, the loosely bound electron is more likely to be ionized, giving higher conversion efficiency for larger values of $a_0$.  Furthermore, the resonant harmonic intensity is relatively higher for lower values of $a_0$ (more bounded states). We explain this in the following way.  For the higher values of $a_0$, the electron is loosely bound to the nucleus. The application of the external dc field may push the same to the continuum, which results in a weak transition from the excited state to the ground state. Hence the lower resonant harmonic intensity is observed for larger values of $a_0$.

\begin{figure}[b]
\centering\includegraphics[totalheight=0.7\columnwidth]{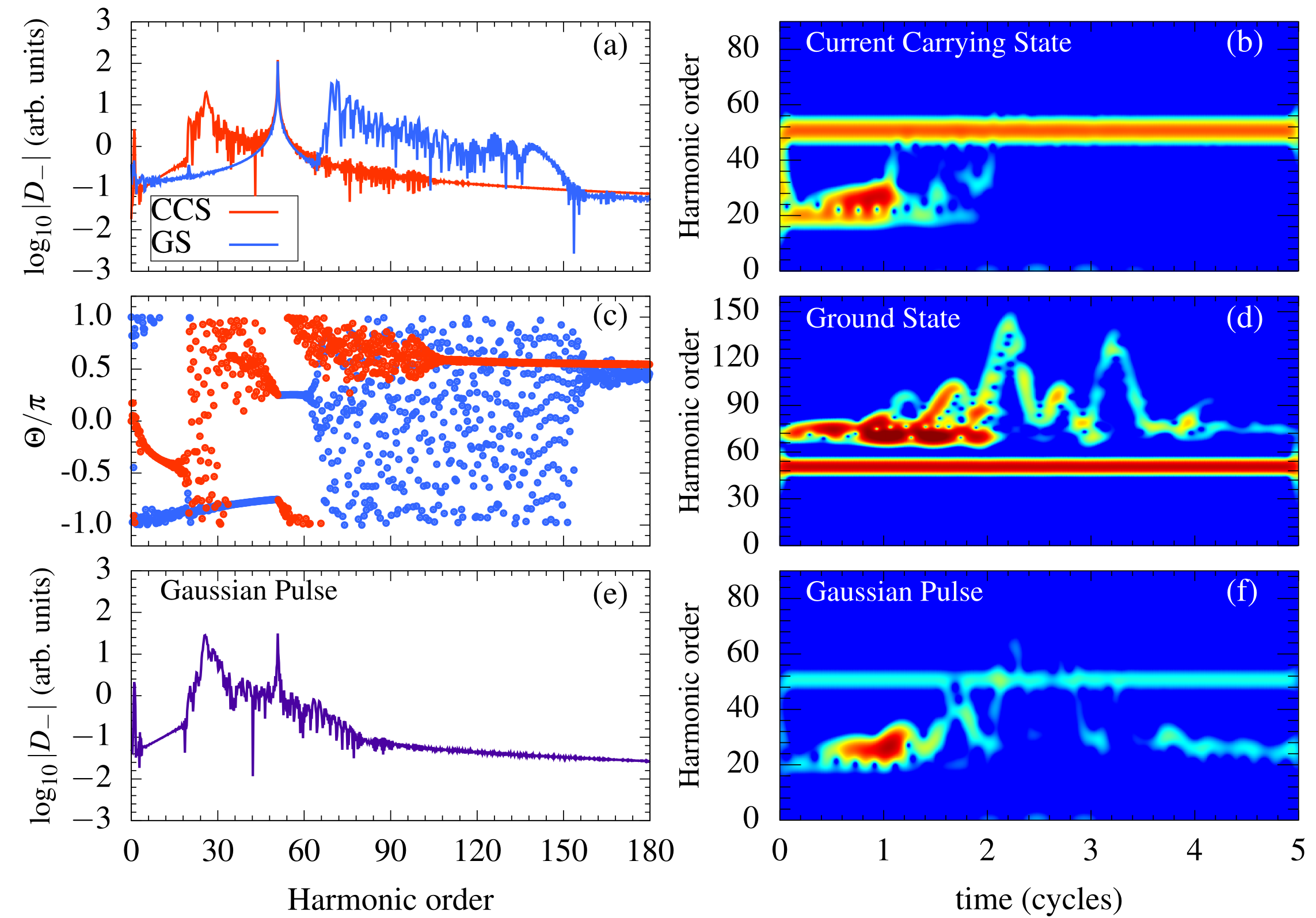}
 \caption{ The current carrying state  (CCS) and ground state (GS) are considered as initial states, and corresponding harmonic spectra (a) are presented, along with the respective phases (c). The potential parameter $a_0 = 1.5$ is used. The time-frequency Gabor transform for CCS (b) and GS (d) are also presented. In (a)-(d) the laser parameters are the same as presented in Fig. \ref{fig2}. The harmonic spectra and the corresponding Gabor transform for CCS using Gaussian driver are presented in (e) and (f) respectively. For (e) and (f), the $\mb{E}(t,3)$ laser pulse is used with peak intensity $I_0 = 5\times 10^{14}$ W cm$^{-2}$ with FWHM duration $T = 2T_0$, $\eta = 2.5T$, and the total duration of $5T_0$.}  
 \label{fig4} 
\end{figure}
 
Recent advances in the field of condensed matter physics made it viable to engineer the materials with precise band-gap using an appropriate number of ions arranged in a specific manner \cite{PhysRevA.103.053121}. Precisely designed nanostructured band-gap material is now routinely used in the context of the higher harmonic generation \cite{Zurr_n_2018,PhysRevA.103.043117}. In this spirit, the model potential as given by Eq. \ref{poten} or any other potential for that matter can be engineered using an adequate arrangement of the ions. The possibility of engineering such model potential, along with the concept of the recollision excitation, makes it feasible to tune the resonant peak as seen in Fig. \ref{fig2}(a). In Fig. \ref{fig3} we present the variation of the ground state ($E_0$) and first excited state energy ($E_1$) with the model potential parameter $a_0$. The resonant excitation in the harmonic spectrum because of the energy difference ($\Delta E/\omega_0 = (E_1 - E_0)/\omega_0$) is also presented in  Fig. \ref{fig3}. It can be further observed from Fig. \ref{fig3} that the resonant harmonic scales very nicely as $\propto 1/\sqrt{a_0}$, which opens the possibility to finely tune the resonant harmonic enhancement by tuning the potential parameter, which we believe is feasible in the near future. 
 
So far, we have propagated the current-carrying state $\ket{\phi_+}$ and studied the resonance enhancement in the harmonic emission equal to the difference between the excited state and the ground state. We investigate the effect of the initial state and the driver envelope in the following way.
In Fig. \ref{fig4}(a), we comapre the harmonic spectra for cases when the initial state is considered to be the ground state $\ket{\phi_0}$, and the current carrying state $\ket{\phi_{p_+}}$ of the  potential with parameter $a_0 = 1.5$. The bound state energies of the $\ket{\phi_0}$ and $\ket{\phi_{p_+}}$ are  respectively found to be $E_0 \sim -4.397$ a.u. and $E_1 \sim -1.504$ a.u. for the potential $V(r,1.5)$ [refer Eq. \ref{poten}]. All other laser parameters are the same as in Fig. \ref{fig2}. It can be observed from the Fig. \ref{fig4}(a) that, when the state $\ket{\phi_{p_+}}$ is propagated, two distinct peaks at $\sim 26^\text{th}$ and $\sim 51^\text{st}$ harmonics are visible. The resonant harmonic at $\sim 51^\text{st}$ is due to the transition from the excited state to the ground state, i.e. $\Delta E/\omega_0 = (E_1-E_0)/\omega_0\sim 51$. However, the peak at $\sim 26^\text{th}$ harmonic corresponds to the ionization threshold of the $\ket{\phi_{p_x}}$, i.e., $E_1 \sim -1.504$ a.u. The harmonic $\lesssim 26$ are severely suppressed in this case because the potential barrier restricts any slow-moving electrons to recombine. Furthermore, when the ground state is propagated, again, we observe a resonant peak at   $\sim 51^\text{st}$ harmonic, resembling the transition from the excited to the ground state. However, another prominent peak is observed   $\sim 77^\text{th}$ harmonic, which is the ionization threshold of   $\ket{\phi_0}$, i.e., the ground state energy $E_0$. Again, because of the potential barrier, the slow-moving electron is stopped, and harmonics $\lesssim 51$ are suppressed. The standard harmonic cutoff as predicted by the \textit{three} step model, i.e. $\sim 3.17 U_p + I_p$, is observed at   $\sim 140^\text{th}$ harmonic, where $U_p = F_0^2/4\omega_0^2 \sim 7.88$  a.u. is the ponderomotive energy of the returning electron and $I_p\sim 4.397$ a.u. is the ionization energy of the ground state. The harmonic phases for both the cases are presented in Fig. \ref{fig4}(c). The phase jump of $\pi$ at the resonance peak $\sim 51^\text{st}$ harmonic is visible. We also observe this phase jump for all the results presented in Fig. \ref{fig2}, which is a clear sign of the resonant phenomenon \cite{PhysRevA.89.053833}. The Gabor transform for both the cases are   presented in Fig. \ref{fig4}(b) and (d). In Fig. \ref{fig4}(b), for the current carrying state propagation, a clear resonant $\sim 51^\text{st}$ harmonic is observed throughout the laser pulse duration, along with a peak $\sim 26^\text{th}$ harmonic. Also, for the ground state propagation, the resonant  $\sim 51^\text{st}$ harmonic is seen.  

 Next, to understand the effect of the laser envelope, we calculated the harmonic spectrum for $\ket{\phi_+}$ state [refer Fig. \ref{fig4}(e)] using the laser pulse $\mb{E}(t,3)$ with $T = 2 T_0$ being the FWHM pulse duration (and $5T_0$ is the total duration), and $\eta = 2.5 T_0$ is used. The profile $\mb{E}(t,3)$ ensures to have the field amplitude $\sim -0.00157$ a.u. at $t = 0$. This non-zero field at $t = 0$ removes the necessity to have the dc field as we used in Fig. \ref{fig2}, which makes the proposed scheme more feasible experimentally. The initial field amplitude can easily be tweaked by varying the envelope phase.
In this case, the resonant harmonic is not that prominent because the initial non-zero field is smaller as compared to the results presented in Fig. \ref{fig4}(a), wherein the non-zero field was $5\%$ of the peak laser field. The intensity of the resonant harmonic depends on the non-zero electric field, as it mimics the Coulombic fields in multielectron configuration and responsible for the dipole transition from the excited state to the ground state.   
 
\section{Concluding remarks}
\label{sec4}

In conclusion, we have proposed a method to control the resonant enhancement in the higher-harmonic generation by tweaking the engineered potential trap, which is experimentally feasible considering the recent advances in atomic and the condensed matter physics. In the presence of the dc electric field, it is found that the electron can have a sustained transition from the excited state to the ground state, giving the enhanced intensity of the harmonic equal to the energy gap between the two levels. We also observed that the laser pulse with the Gaussian envelope and proper phase could also have a non-zero electric field at the beginning, triggering the resonant enhancement. The position of the resonant harmonic in the harmonic spectrum is found to be the property of the potential used in the Hamiltonian. For a model potential used in the study, the resonant harmonic is observed to follow the $\propto 1/\sqrt{a_0}$ scaling. We like to emphasize that the proposed method is not restricted to the model potential used in the study; however, it will work for other potentials and support the bound states.  
  
\section*{Acknowledgments} Authors would like to acknowledge the DST-SERB, Government of India, for funding the project CRG/2020/001020.


%

\end{document}